\documentclass[aps,pre,twocolumn,showpacs,superscriptaddress]{revtex4}
\usepackage{graphicx,color,epsfig}
\usepackage{amssymb,amsmath,amsfonts}
\begin{document}

\title{Collective behavior of chaotic oscillators  with environmental coupling}
\author{C. Quintero-Quiroz}
\affiliation{Departament de F\'isica i Enginyeria Nuclear, Universitat Polit\'ecnica de Catalunya, 08222 Terrassa, 
Barcelona, Spain}
\author{M. G. Cosenza}
\affiliation{Grupo de Caos y Sistemas Complejos, 
Centro de F\'isica Fundamental, 
Universidad de Los Andes, M\'erida, M\'erida 5251, Venezuela}

\begin{abstract}
We investigate the collective behavior of a system of chaotic R\"ossler oscillators 
indirectly coupled through a common environment that possesses its own dynamics 
and which in turn is modulated by the interaction with the oscillators. By
varying the parameter representing the coupling strength between the
oscillators and the environment, we find two collective states previously not reported in systems with environmental coupling:
(i) nontrivial collective behavior, characterized by 
a periodic evolution of macroscopic variables coexisting with the
local chaotic dynamics; and 
(ii) dynamical clustering, consisting of the formation of differentiated subsets of synchronized elements 
within the system. These states are relevant for many physical and biological systems
where interactions with a dynamical environment are frequent. 
\end{abstract}


\maketitle

Many physical, biological, and social systems exhibit global interactions;
i. e.,  all the elements in the system are subject to a common influence. 
These systems have been widely studied in many theoretical and experimental models
\cite{Kuramoto,Nakagawa,Hadley,Gruner,Wie,Kaneko2,Yakovenko,Media,Ojalvo,Wang,DeMonte,Taylor,Roy}.
A global interaction may consist of an external field acting on the elements,
as in a driven (or unidirectionally coupled) dynamical system; or it may originate from
the mutual interactions between the elements, in which case, we refer to an autonomous dynamical system. 
Recently, there has been interest in the investigation of systems of dynamical elements subject to a global interaction
through a common environment or medium that possesses its own dynamics. 
In this case, the state of each element in the system influences the
environment, and the state of the environment in turn affects
the elements. This type of global interaction has been denominated as environmental coupling \cite{Katriel,Resmi,Sharma,Kurths}. 
Examples of such systems include chemical and genetic oscillators where coupling is through
exchange of chemicals with the surrounding medium \cite{Toth,Kustnesov,Wang2},
ensembles of cold atoms interacting with a coherent electromagnetic field \cite{Java}, 
and coupled circadian oscillators due to common global neurotransmitter oscillation \cite{Gonze}. 
Since the elements are not directly interacting with each other but through a common medium, this
configuration has also been called indirect coupling \cite{Chinos,Amit}, or relay coupling \cite{Feudel}.

Most models of systems subject to environmental coupling have mainly focused on the study of the synchronization behavior of two
oscillators interacting with a dynamical element. In this paper we investigate the collective behavior arising in a system 
consisting of many chaotic oscillators subject to environmental coupling.  
The large number of oscillators allows the emergence of
collective states not present in those previous models. 
By varying the coupling strength between the chaotic 
oscillators and the environment, we find these collective states:
(i) nontrivial collective behavior,  i. e., non-statistical fluctuations in the mean-field of the ensemble,  manifested by 
a periodic evolution of macroscopic variables coexisting with the
local chaotic dynamics  \cite{Kaneko3,Chate}; and 
(ii) dynamical clustering, i. e., the formation of differentiated subsets of synchronized elements 
within the system \cite{Kaneko4}.

We consider a system of $N$ chaotic R\"ossler oscillators coupled through a common environment
that can receive feedback from the system, 
\begin{eqnarray}
 \label{Sys}
 \dot x_i &=& -y_i -z_i +\varepsilon_2  w , \nonumber  \\
 \dot y_i &=&  x_i + a y_i,   \\  
 \dot z_i &=&  b + z_i(x_i- c)  , \nonumber \\
 \dot w   &=& -\lambda   w + \frac{\varepsilon_1}{N} \sum_{j=1}^N x_j \,  ,
 \label{Envi}
\end{eqnarray}
where $x_i(t), y_i(t), z_i(t)$ describe the state variables of oscillator $i=1,2,\ldots,N$, at time $t$; 
$w(t)$ represents the state of the environment at $t$; $a,b,c$ are parameters of the local dynamics; the parameter 
$\varepsilon_2$ measures the strength of the global influence from the environment to the oscillators; and
$\varepsilon_1$ represents the intensity of the global feedback to the environment. The damping parameter $\lambda$ characterizes 
the intrinsic dynamics of the environment, which decays in time in absence of feedback from the oscillators.
The form of the global coupling in the system Eqs.~(\ref{Sys})-(\ref{Envi}) is non-diffusive.

A synchronized state in the system at time $t$ corresponds to $x_i(t)=x_j(t)$, $y_i(t)=y_j(t)$, $z_i(t)=z_j(t)$, $\forall i, j$. 
This condition has also been denoted as amplitude synchronization. 
The occurrence of stable synchronization in the system Eq.~(\ref{Sys}) can be numerically characterized 
by the asymptotic time-average $\langle \sigma \rangle$ of the instantaneous standard deviations
of the distribution of state variables, defined as 
\begin{eqnarray}
\langle \sigma \rangle &=&  \frac{1}{T-\tau} \sum_{t=\tau}^T \sigma (t) , \\
\sigma (t)  &=&   \left[  \frac{1}{N} \sum_{i=1}^N   (x_i- \bar X)^2 + (y_i- \bar Y)^2 + (z_i- \bar Z)^2 \right]^{1/2}  
\end{eqnarray}
where $\tau$ is a discarded transient time, and the mean values are defined as  
\begin{eqnarray}
\bar{X}(t) &=&  \frac{1}{N} \sum_{j=1}^N x_j(t) , \\
\bar{Y}(t)   &=&   \frac{1}{N} \sum_{j=1}^N y_j(t) ,\\
\bar{Z}(t)&=&  \frac{1}{N} \sum_{j=1}^N z_j(t)  .
\end{eqnarray}
Then, a synchronization state corresponds to a value $\langle \sigma \rangle=0$. 
On the other hand, the instantaneous phase of the trajectory of oscillator $i$ projected on the plane $(x_i,y_i)$ can be defined as
\begin{equation}
\phi_i(t) = \tan^{-1} \left( \frac{y_i(t)}{x_i(t)} \right). 
\end{equation}

To characterize a collective state of phase synchronization on the plane $(x,y)$, we calculate the
asymptotic time-average  quantity
{\small
\begin{equation}
\Phi = \frac{1}{T-\tau} \sum_{t=\tau}^T \left[\left(\frac{1}{N} \sum_{j=1}^N \sin \phi_i(t) \right)^2 
+  \left(\frac{1}{N} \sum_{j=1}^N \cos \phi_i(t)\right)^2 \right] .
\end{equation}
}
Then, a collective phase synchronization state corresponds to a value $\Phi=1$. 

We have fixed the local parameters in Eqs.~(\ref{Sys})-(\ref{Envi}) at values the $a=b=0.1$ and $c=18$, 
for which a R\"ossler 
oscillator displays chaotic behavior. The damping parameter for the environment is fixed at the value $\lambda=1$.
Then, we numerically integrate the system Eqs.~(\ref{Sys})-(\ref{Envi}) with given size $N$ 
for different values of the coupling parameters $\varepsilon_1$ and $\varepsilon_2$.
We employ a fourth-order Runge-Kutta scheme with fixed integration step $h=0.01$. The initial conditions for the variables 
$x_i, y_i$ 
were randomly distributed with uniform probability on the interval $[-20, 20]$, and those for the variables $z_i$ on the interval 
$[0,5]$, $\forall i$. 

%%%%%%%%%%%%%%%%%%%%%%%%%%%%%%%%%%%%%%%%%%%%%%%%%%%%%%%%%%%%%%%%%%%%%%
\begin{figure}[h]
\begin{center}
\includegraphics[width=0.96\linewidth,angle=0]{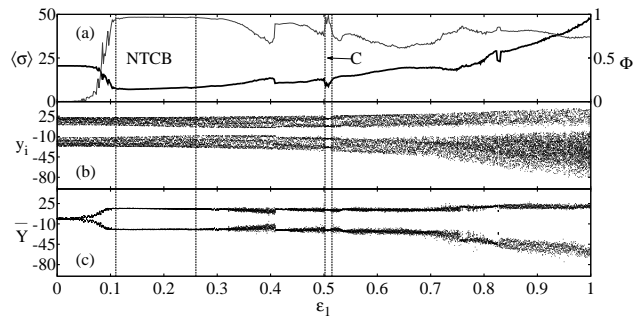}
\end{center}
\caption{(a) The quantities $\langle \sigma \rangle$ (thick line, left vertical axis) and $\Phi$ (thin line, right vertical axis) 
as functions of the coupling strength $\varepsilon_1$, for the system Eqs.~(\ref{Sys})-(\ref{Envi}) with $\varepsilon_1=\varepsilon_2$. 
The labels NTCB and C indicate the regions of the coupling where nontrivial collective behavior and dynamical clustering occur, respectively.  
Fixed parameters values are:  $a=b=0.1, c=18, \lambda=1, N=10^3, \tau=10^3, T=7 \times 10^3$. 
(b) Bifurcation diagram of the values $y_i$, when $x_i=0$ on the plane $(x_i,y_i)$, for one oscillator as a function of $\varepsilon_1$.
(c) Bifurcation diagram of the component $\bar Y$ of the mean field, when $\bar X=0$ on the plane $(\bar X, \bar Y)$,
as a function of $\varepsilon_1$.
For each value of $\varepsilon_1$, $300$ consecutive values of $y_i$ and $\bar Y$ have been plotted in (b) and (c), 
after discarding the transient time $\tau$. 
}
\label{F1}      
\end{figure}
%%%%%%%%%%%%%%%%%%%%%%%%%%%%%%%%%%%%%%%%%%%%%%%%%%%%%%%%%%%%%%%%%%%%%%

Figure~(\ref{F1}a) shows the statistical quantities $\langle \sigma \rangle$ and $\Phi$ as functions of $\varepsilon_1$ for 
the system Eqs.~(\ref{Sys})-(\ref{Envi}), with $\varepsilon_1=\varepsilon_2$. The environmental coupling induces some appreciable degree 
of both forms of synchronization, amplitude ($\langle \sigma \rangle$ small) and phase ($\Phi \to 1$), in the range of parameter 
$\varepsilon_1 \in[ 0.11, 0.26]$. However, for larger values of 
the coupling strength, $\varepsilon_1 >0.9$, these two synchronization measures do not behave in the same fashion:
the state variables are quite disperse ($\langle \sigma \rangle$ large) while the phases are still close to each other ($\Phi \simeq 0.8$).

To analyze the dynamical behavior of the system at both the local and the global levels of description, 
we consider the projections of the trajectories of one oscillator and that of the mean field of the system on the planes $(x_i,y_i)$ and
$(\bar X, \bar Y)$, respectively. Then, in Fig.~(\ref{F1}b) we plot the bifurcation diagram of the values $y_i$ when $x_i=0$,
as a function of $\varepsilon_1$. The two chaotic bands mainly observed as the coupling parameter is varied reflect the typical one-scroll
structure of the projected local R\"ossler attractor. However, there is a range of $\varepsilon_1$ where a distinguishable periodic window 
(period-four behavior) emerges in the local dynamics of the coupled oscillator. Similarly, in Fig.~(\ref{F1}c) we show the bifurcation diagram
of the component $\bar Y$ of the mean field for $\bar X=0$, as a function of $\varepsilon_1$. The mean field unveils the presence
of a window of global period-two behavior for $\varepsilon \in [0.11, 0.26]$, where there is an increase in the amount of both forms of 
synchronization. Thus, in this region of the coupling parameter, collective periodic motion coexists with chaos at the local level,
indicating the occurrence of nontrivial collective behavior. In this representation, collective periodic states at a given value of the coupling
$\varepsilon_1$ appear in $\bar Y$ as sets of short vertical segments which correspond to intrinsic fluctuations of the periodic orbit
of the mean field. At a value $\varepsilon_1=0.05$ a pitchfork bifurcation in the dynamical behavior of $\bar Y$ takes place, 
from a statistical fixed point to a global period-two state, where the time series of $\bar Y$ alternately  moves between the
corresponding neighborhoods of two separated, well-defined values. 

Figure~(\ref{F1}) reveals two relevant behaviors in different ranges of the coupling parameter: 
(i) a nontrivial collective behavior; and (ii) a periodic, desynchronized motion of the local dynamics.

%%%%%%%%%%%%%%%%%%%%%%%%%%%%%%%%%%%%%%%%%%%%%%%%%%%%%%%%%%%%%%%%%%%%%%
\begin{figure}[h]
\begin{center}
\includegraphics[width=0.45\linewidth,angle=0]{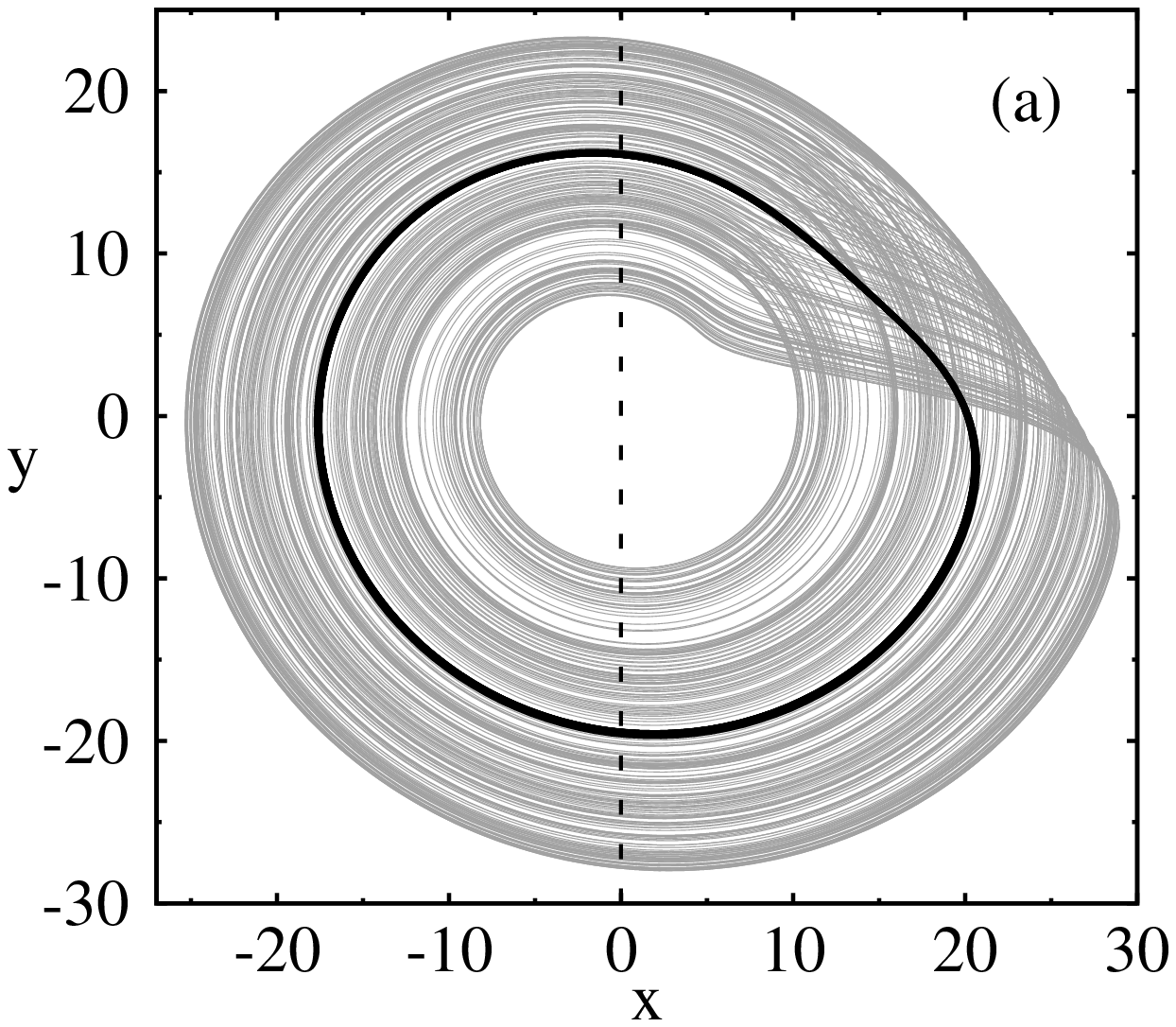}
\hspace{6mm}
\includegraphics[width=0.45\linewidth,angle=0]{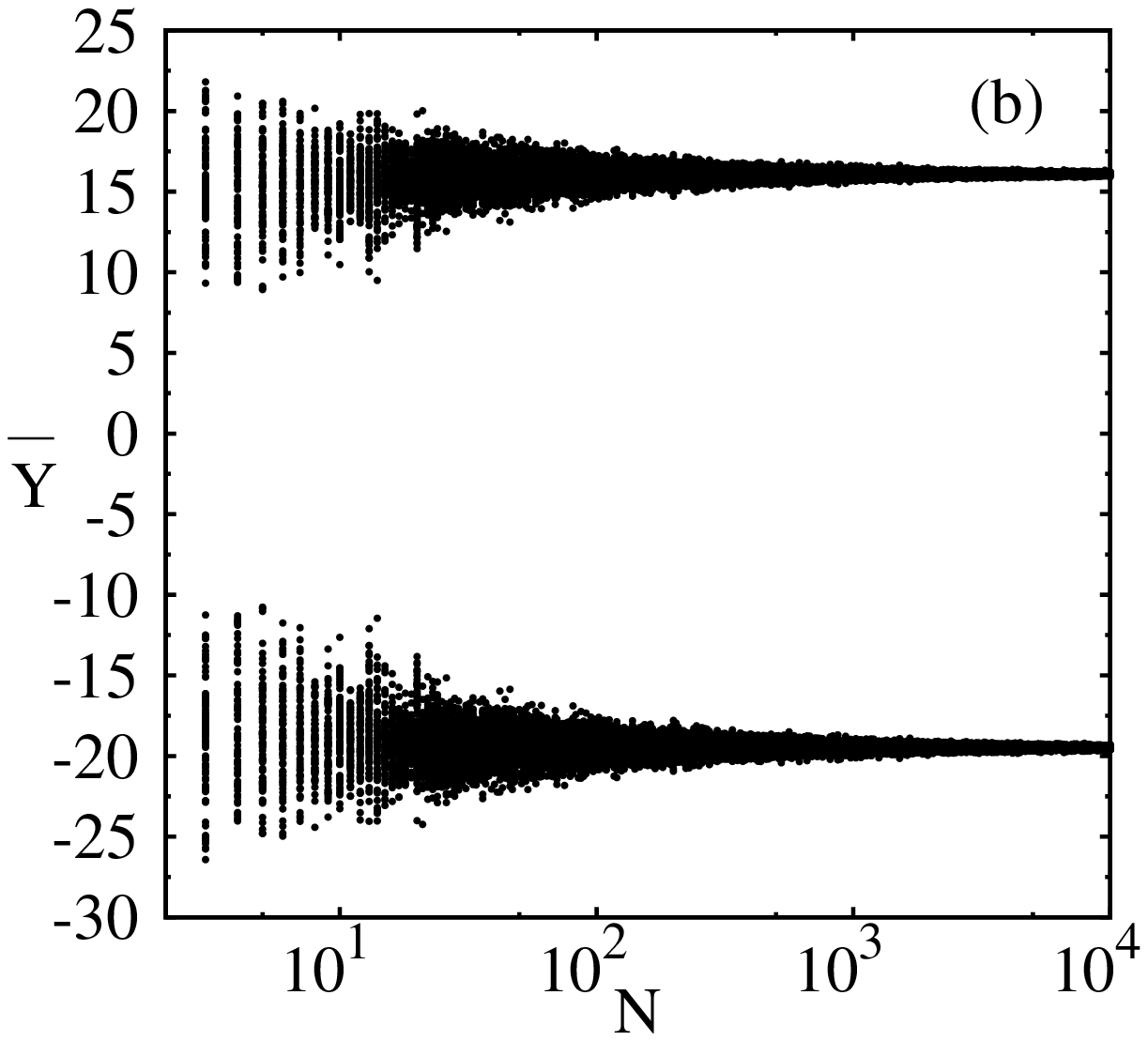}
\end{center}
\caption{(a) Projections on the plane $(x,y)$ of the trajectories corresponding to one oscillator (gray line) and to the mean field (black line) 
for the system Eqs.~(\ref{Sys})-(\ref{Envi}) with $N=1000$, $\varepsilon_1=\varepsilon_2=0.2$.
(b) Component $\bar Y$  of the  mean field when $\bar X=0$, corresponding to the  dashed line in (a), as a function of 
the system size $N$ (in log scale). 
}
\label{F2}     
\end{figure}
%%%%%%%%%%%%%%%%%%%%%%%%%%%%%%%%%%%%%%%%%%%%%%%%%%%%%%%%%%%%%%%%%%%%%%

In order to clarify the nature of behavior (i), we show in Fig.~(\ref{F2}a) a superposition of the projections on the plane $(x,y)$ 
of the trajectories corresponding to one oscillator and to the mean field for the system Eqs.~(\ref{Sys})-(\ref{Envi}), respectively.
We observe that the trajectory associated to the oscillator is chaotic, while that corresponding to the mean field of the system is
periodic. The trajectories of all the oscillators are not synchronized; they move closely together, displaying some dispersion, 
in analogy to the motion of a swarm of insects. This dispersion is manifested in the width of the periodic orbit of the mean field.
 Figure(\ref{F2}b) shows the  
segments that constitute the component $\bar Y$ when  $\bar X=0$ for the periodic orbit of the mean field, 
as a function of the system size $N$. The width of the segments shrinks as $N$ increases, according to the law of large numbers,
indicating that the periodic orbit of the mean field becomes better defined in the large system limit. 
Thus, when the size of the system is increased, the width of the periodic orbit
of the mean field decreases, but its amplitude does not change in Fig.~(\ref{F2}a).
This is a phenomenon of nontrivial collective behavior
induced by the environment. 

\begin{figure}[h]
\begin{center}
\includegraphics[width=0.45\linewidth,angle=0]{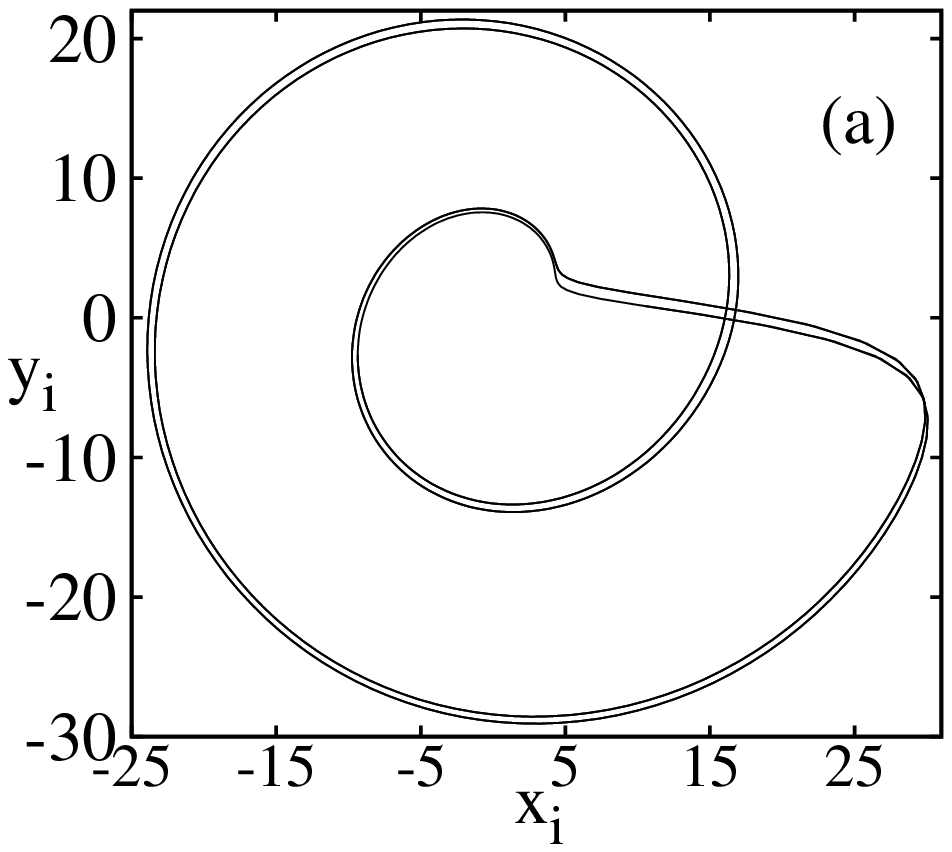}
\hspace{6mm}
\includegraphics[width=0.45\linewidth,angle=0]{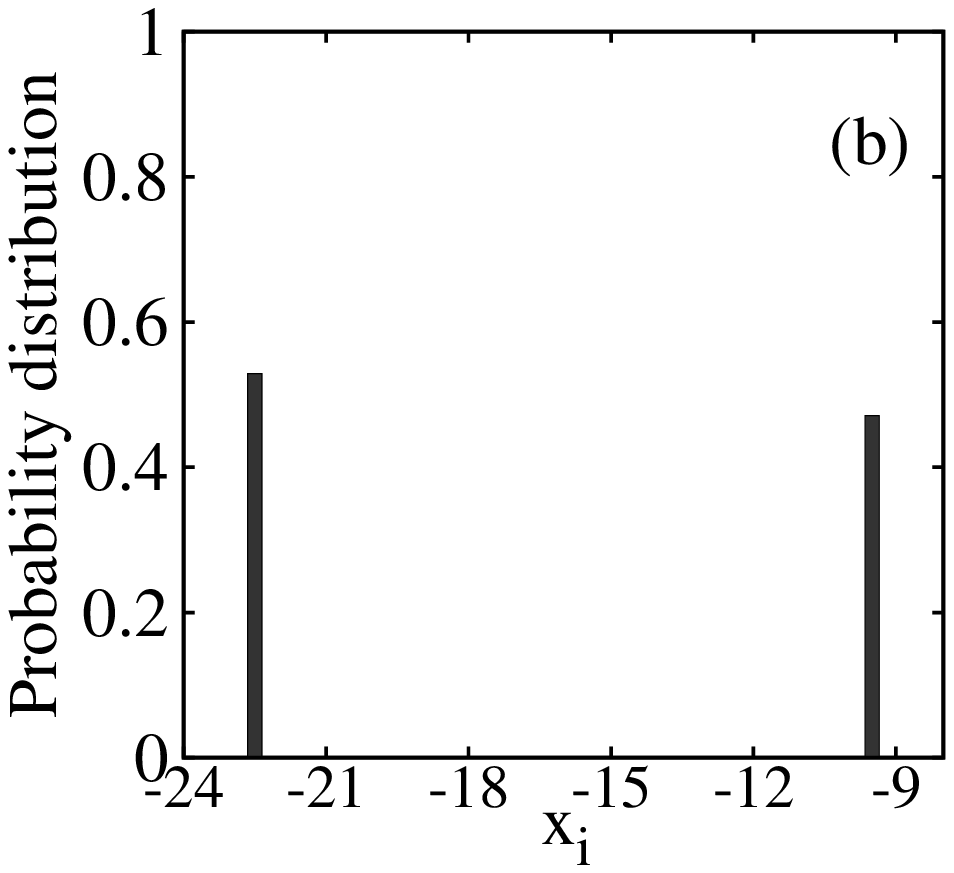}
\end{center}
\caption{(a) Projection on the plane $(x_i,y_i)$ of the trajectory of one oscillator in the system Eqs.~(\ref{Sys})-(\ref{Envi}), with
$\varepsilon_1=\varepsilon_2=0.516$, $N=10^4$.
(b) Probability distribution of the $x_i$ component of the state variables of the $N$ oscillators in the system at $t=3 \times 10^3$. 
}
\label{F3}
\end{figure}

To elucidate the observed periodic behavior (ii), Fig.~(\ref{F3}a) shows a projection on the plane $(x_i,y_i)$ of the trajectory 
of one oscillator in system Eqs.~(\ref{Sys})-(\ref{Envi})
for a coupling parameter value $\varepsilon_1=0.516$, within the periodic-four window in Fig.~(\ref{F1}b).
The continuous trajectory of the oscillator corresponds to a period-two orbit, manifested as a period-four orbit in 
the discrete time series of $y_i$ when taking the Poincar\`e section at $x_i=0$. This periodic behavior in the local dynamics 
is induced by the environmental coupling in this range of parameters. However, as seen in Fig.~(\ref{F1}a), the periodic motion of the oscillators
in this window of the coupling parameter is not completely synchronized. Figure~(\ref{F3}b) shows the probability distribution of the $x_i$ 
component of the state variables of the $N$ oscillators in the system at a given time. We observe that the oscillators become segregated into
two groups or clusters of comparable sizes; each group displaying a synchronized period-four orbit, 
but not synchronized to the other group. This a phenomenon of dynamical clustering or cluster synchronization.

%%%%%%%%%%%%%%%%%%%%%%%%%%%%%%%%%%%%%%%%%%%%%%%%%%%%%%%%%%%%%%%%%%%%%%
%%%%%%%%%%%%%%%%%%%%%%%%%%%%%%%%%%%%%%%%%%%%%%%%%%%%%%%%%%%%%%%%%%%%%%

We recall that nontrivial collective behavior has been observed in autonomous systems with mean field global coupling 
with discrete time maps \cite{Kaneko3,Chate,PTP}, as well as with continuous time flows \cite{Pikovsky}, as local chaotic dynamics. Similarly,
dynamical clustering commonly occurs in autonomous globally coupled chaotic systems with either discrete \cite{Kaneko4,We} or continuous time 
\cite{Zanette} dynamics.
Our result shows that both phenomena can also occur in chaotic systems subject to non-diffusive environmental coupling.

\begin{figure}[h]
\begin{center}
\includegraphics[width=0.7\linewidth,angle=0]{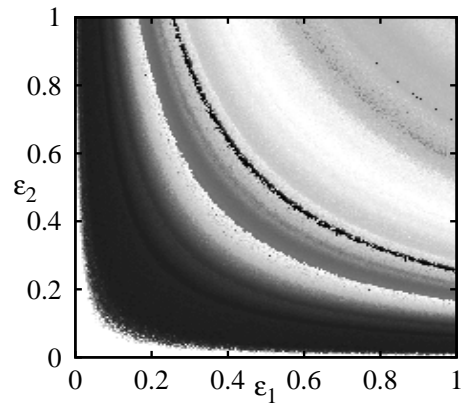}
\end{center}
\caption{Phase synchronization measure $\Phi$ on the space of parameters $(\varepsilon_1, \varepsilon_2)$ for 
the system Eqs.~(\ref{Sys})-(\ref{Envi}). The values of $\Phi$ are indicated by
shades of gray, from white ($\Phi=0$) to black ($\Phi=1$, full phase synchronization). 
}
\label{F4}
\end{figure}

Figure~(\ref{F4}) shows the phase synchronization measure $\Phi$ for the oscillators in the system Eqs.~(\ref{Sys})-(\ref{Envi}),
calculated on the space of the coupling parameters $(\varepsilon_1, \varepsilon_2)$. The phase diagram is symmetric about the diagonal 
$\varepsilon_1=\varepsilon_2$. The regions of parameters where  the value of $\Phi$ is large correspond to the main collective behaviors
observed in the system, i. e., nontrivial collective behavior and dynamical clustering: they constitute two different dynamical manifestations of 
phase synchronization states. 

%%%%%%%%%%%%%%%%%%%%%%%%%%%%%%%%%%%%%%%%%%%%%%%%%%%%%%%%%%%%%%%%%%%%%%
%%%%%%%%%%%%%%%%%%%%%%%%%%%%%%%%%%%%%%%%%%%%%%%%%%%%%%%%%%%%%%%%%%%%%%

In summary, we have investigated the collective behavior of a system of $N$ chaotic oscillators subject 
to environmental coupling. Previous works  have mostly considered 
a number of $N=2$ oscillators in systems with environmental or indirect coupling. 
The large number of oscillators that we have employed permits the occurrence of
collective states not present in those previous models, i. e.,  nontrivial collective behavior and dynamical clustering. 
We have verified that these collective states also arise for other forms of the local chaotic dynamics in systems with environmental coupling.
Clustering and nontrivial collective behavior have been suggested as possible mechanisms for cell differentiation 
and self-organization in complex systems \cite{Kaneko5}.
Thus, our results become relevant for many biological
systems that can be described as populations of oscillators interacting with a common dynamical environment.

\subsection*{Acknowledgments}
This work is supported by project No. C-1827-13-05-B from CDCHTA, 
Universidad de Los Andes, M\'erida, Venezuela. M. G. C. is grateful to the Senior Associates Program of 
the Abdus Salam International Centre for Theoretical Physics, Trieste, Italy, for the visiting opportunities.


\begin{thebibliography}{99}
\bibitem{Kuramoto} Y. Kuramoto, \textit{Chemical Oscillations, Waves and Turbulence} (Springer, Berlin, 1984).  
\bibitem{Nakagawa} N. Nakagawa, Y. Kuramoto, Physica D \textbf{75},(1994) 74-80.
\bibitem{Hadley} K. Wiesenfeld, P. Hadley, Phys. Rev. Lett. \textbf{62}, (1989) 1335-1338. 
\bibitem{Gruner} G. Gr\"uner, Rev. Mod. Phys. \textbf{60}, (1988) 1129-1181.
\bibitem{Wie} K. Wiesenfeld, C. Bracikowski, G. James, and R. Roy, Phys. Rev. Lett. \textbf{65}, (1990) 1749-1752.
\bibitem{Kaneko2} K. Kaneko, Physica D \textbf{41}, (1990)  137-172.
\bibitem{Yakovenko} V. M. Yakovenko,  in \textit{Encyclopedia of Complexity and System Science}, edited by R. A. Meyers (Springer, New York, 2009).
\bibitem{Media} J. C. Gonz\'alez-Avella, V. M. Eguiluz, M. G. Cosenza, K. Klemm, J. L. Herrera, M. San Miguel,
Phys. Rev. E \textbf{73}, (2006) 046119.
\bibitem{Ojalvo} J. Garcia-Ojalvo, M. B. Elowitz,  S. H. Strogatz, Proc. Natl. Acad. Sci. U.S.A. \textbf{101}, (2004) 10955-10960.
\bibitem{Wang} W. Wang, I. Z. Kiss, J. L. Hudson, Chaos \textbf{10}, (2000) 248-256.
\bibitem{DeMonte} S. De Monte, F. d’Ovidio, S. Dan\o{},  P. G. S\o{}rensen, Proc. Natl. Acad. Sci. U.S.A. \textbf{104}, (2007) 18377-18381. 
\bibitem{Taylor} A. F. Taylor,  M. R. Tinsley, F. Wang, Z. Huang,  K. Showalter, Science \textbf{323}, (2009) 614-617.
\bibitem{Roy} A. M. Hagerstrom, T. E. Murphy, R. Roy, P. H\"ovel, I. Omelchenko,  E. Sch\"oll, Nature Phys. \textbf{8}, (2012) 658-661.
\bibitem{Katriel} G. Katriel, Physica D \textbf{237}, (2008) 2933-2944.
\bibitem{Resmi} V. Resmi, G. Ambika, R. E. Amritkar, Phys. Rev. E \textbf{81}, (2010) 046216.
\bibitem{Sharma} A. Sharma, M. D. Shrimali, S. K. Dana, Chaos \textbf{22}, 023147 (2012).
\bibitem{Kurths} R. Suresh, K. Srinivasan, D. V. Senthilkumar, K. Murali, M. Lakshmanan, J. Kurths, arXiv:1304.1254v2 (2014).
\bibitem{Toth} R. Toth, A .F. Taylor, M. R. Tinsley, J. Phys. Chem. B \textbf{110} (2006) 10170-10176.
\bibitem{Kustnesov} A. Kuznetsov, M. Kaern, N. Kopell, SIAM J. Appl. Math. \textbf{65}, (2005) 392-425.
\bibitem{Wang2} R. Wang, L. Chen, J. Biol. Rhythms \textbf{20} (2005) 257-269.
\bibitem{Java} J. Javaloyes, M. Perrin, A. Politi, Phys. Rev. E \textbf{78}, (2008) 011108.
\bibitem{Gonze} D. Gonze, S. Bernard, C. Waltermann, A. Kramer, H. Herzel,  Biophys. J. \textbf{89} (2005) 120-129.
\bibitem{Chinos} B. W. Li, C. Fu, H. Zhang, X. Wang, Phys. Rev. E \textbf{86}, 046207 (2012).
\bibitem{Amit} A. Sharma,  M. D. Shrimali, Pramana \textbf{77}, (2011) 881-889.
\bibitem{Feudel} R. Guti\'errez, R. Sevilla-Escoboza, P. Piedrahita, C. Finke, U. Feudel, J. M Buld\'u, 
G. Huerta-Cuellar, R. Jaimes-Re\'ategui, Y. Moreno, S. Boccaletti, Phys. Rev. E  \textbf{88}, (2013) 052908.
\bibitem{Kaneko3} K. Kaneko,  Phys. Rev. Lett. \textbf{65}, (1990) 1391-1394.
\bibitem{Chate} H. Chat\'e, P. Manneville, Prog. Theor. Phys. \textbf{87}, (1992) 1-60.
\bibitem{Kaneko4} K. Kaneko, Physica D \textbf{41}, (1990) 137-172.
\bibitem{PTP} M. G.  Cosenza, J. Gonz\'alez, Prog. Theor. Phys. \textbf{100}, (1998) 21-38.
\bibitem{Pikovsky} A. S. Pikovsky, M. G. Rosenblum, J. Kurths, Europhys. Lett. \textbf{34}, (1996) 165-170.
\bibitem{We} M. G. Cosenza, A. Parravano, Phys. Rev. E \textbf{64}, (2001) 036224.
\bibitem{Zanette} D. H. Zanette, A. S. Mikhailov, Phys. Rev. E \textbf{57} (1998) 276-281.
\bibitem{Kaneko5}  K. Kaneko, I. Tsuda, \textit{Complex Systems: Chaos and Beyond}, (Springer, Berlin, 2000).
 \end{thebibliography}
\end{document}